\documentclass[aps,prl,nofootinbib,reprint,superscriptaddress]{revtex4-1}
\usepackage{amsmath}
\usepackage{graphicx}
\usepackage{epsfig}
\usepackage{amssymb}
\usepackage{enumitem}

\def\be{\begin{equation}}
\def\ee{\end{equation}}
\def\bea{\begin{eqnarray}}
\def\eea{\end{eqnarray}}

\def\diff{\mathrm{d}}
\def\dd{\mathrm{d}}
\def\ex{\mathrm{e}}
\def\me{\mathrm{e}}
\def\ii{\mathrm{i}}
\def\nn{\nonumber}
\def\wzero{w}
\def\wone{u}
\def\phizero{a}
\def\Esusy{E_{\mathrm{susy}}}
\def\bone{b_1}
\def\btwo{b_2}
\def\a{\mathtt{a}}
\def\c{\mathtt{c}}
\def\R{\mathbb{R}}
\def\F{\mathcal{F}}
\def\A{\mathcal{A}}
\def\deltawzero{\delta\wzero}
\def\deltawone{\delta\wone}
\def\cpsi{\gamma}
\def\Ssusy{S_{\mathrm{susy}}}
\def\C{\mathcal{C}}
\def\deltanew{\varsigma'}
\def\tildedelta{\varsigma'}
\def\gammanew{\varsigma}
\def\Deltanew{\Box}
\def\tx{x}
\def\tr{\rho}
\def\rhoc{\epsilon}
\def\newFtwo{k_1}
\def\newhtwo{k_2}
\def\newhfour{k_3}
\def\newhsix{k_4}
\def\Stot{S}

\begin{document}

\title{The holographic supersymmetric Casimir energy}
\author{Pietro Benetti Genolini}
\affiliation{Mathematical Institute, University of Oxford, Woodstock Road, Oxford OX2 6GG, U.K.}
\author{Davide Cassani}
\affiliation{LPTHE, Sorbonne Universit\'es UPMC Paris 06 and CNRS, UMR 7589, F-75005, Paris, France}
\author{Dario Martelli}
\affiliation{Department of Mathematics, King's College London, The Strand, London, WC2R 2LS, U.K.}
\author{James Sparks}
\affiliation{Mathematical Institute, University of Oxford, Woodstock Road, Oxford OX2 6GG, U.K.}

\begin{abstract}
\noindent 
We consider a general class of asymptotically locally AdS$_5$ solutions of minimal gauged supergravity, which are dual to superconformal field theories on curved backgrounds $S^1\times M_3$ preserving two supercharges. We demonstrate that standard holographic renormalization corresponds to a scheme that breaks supersymmetry. We propose new boundary terms that restore supersymmetry, and show that for smooth solutions with topology $S^1\times \R^4$ the improved on-shell action reproduces both the supersymmetric Casimir energy and the field theory BPS relation between charges.

\end{abstract}

\maketitle

\section{I. The supersymmetric Casimir energy}\label{sec1}

In \cite{Assel:2014paa, Assel:2015nca} a new observable of $d=4$ superconformal field theories has been introduced: 
the supersymmetric Casimir energy. This is defined by putting the  theory on 
certain curved backgrounds 
$M_4=S^1_\beta\times M_3$, where $S^1_\beta$ is a circle of length $\beta$ and 
$M_3$ is a compact three-manifold. These are rigid supersymmetric backgrounds, 
and the supersymmetric Casimir energy is defined as
\bea\label{Esusy}
\Esusy &=& -\lim_{\beta\rightarrow \infty}\frac{\diff}{\diff\beta} \log Z^{\mathrm{susy}}_{S^1_\beta\times M_3}~.
\eea
Here the partition function $Z^{\mathrm{susy}}$ is computed with  
periodic boundary conditions 
for the fermions around $S^1_\beta$. A key point is that, unlike the vacuum energy of general $d=4$
conformal field theories (CFTs), $\Esusy$ is scheme-independent and thus an intrinsic observable.

The rigid supersymmetric backgrounds of interest comprise a metric on $M_4$ of the form
\bea\label{M4}
g_{4} \,&=&\, \diff \tau^2 + g_3  \,=\, \diff\tau^2 + (\diff \psi + \phizero)^2 + 4\ex^{\wzero} \diff z\diff\bar{z}\ , \ \ 
\eea
where $\tau\sim\tau+\beta$ is a coordinate on $S^1_\beta$. 
The vector $\partial_\psi$ is Killing, and generates a transversely 
holomorphic foliation of $M_3$, with local transverse complex coordinate $z$.
The local one-form $\phizero$ satisfies
$\diff\phizero = \ii\wone\, \ex^{\wzero} \diff z\wedge \diff\bar{z}$, 
where $\wzero=\wzero(z,\bar{z})$, $\wone=\wone(z,\bar{z})$. 
In addition there is a non-dynamical Abelian gauge field, 
which couples to the R-symmetry current and 
arises when the field theory is coupled to background conformal supergravity, given by
\bea\label{A}
A &=& \tfrac{\ii}{8}\wone\, \diff\tau+
\tfrac{1}{4}\wone(\diff\psi+\phizero)+\tfrac{\ii}{4}(\partial_{\bar{z}}\wzero\, \diff\bar{z}-\partial_z\wzero\, \diff z)\nn\\
&&\,\ \qquad +\cpsi\, \diff\psi+\diff\lambda(z,\bar{z})~.
\eea
Notice that the second line is locally pure gauge; however, the constant $\cpsi$ will play an important role. 

The background geometry thus depends on the choice of the two functions $\wzero(z,\bar{z})$, $\wone(z,\bar{z})$, and via~(\ref{Esusy}) the supersymmetric Casimir energy  also {\it a priori} depends on this choice. 
These backgrounds admit two supercharges of opposite R-charge, and associated to each of these
is an integrable complex structure (\emph{i.e.}\ they are  ambi-Hermitian). 
In \cite{Closset:2013vra} it is argued that the supersymmetric partition function depends 
on the background only via the choice of complex structure(s). In the present 
set-up, this implies that $Z^{\mathrm{susy}}$ depends only 
on the transversely holomorphic foliation generated by $\partial_\psi$. 
In particular, deformations of $\wzero(z,\bar{z})$ and $\wone(z,\bar{z})$ that leave this foliation fixed 
should not change~$\Esusy$.

Later in this paper we will focus on the case that topologically $M_3\cong S^3$. Here we may 
embed $S^3\subset\R^4=\R^2\oplus\R^2$, and write\footnote{In this paper our conventions are such that $b_1,b_2>0$.}
$\partial_\psi = \bone \partial_{\varphi_1}+\btwo
\partial_{\varphi_2}$, where $\varphi_1$, $\varphi_2$ are standard $2\pi$ 
periodic azimuthal angles. In this case the above statements imply 
that $\Esusy$ should depend only on $b_1$, $b_2$, and the explicit  calculation in \cite{Assel:2014paa} gives
\bea\label{Esusyb1b2}
\Esusy &=& \frac{2(\bone+\btwo)^3}{27\bone\btwo}(3\c-2\a) + \frac{2}{3}(\bone+\btwo)(\a- \c)\,  .\ \  
\eea
Here $\a$ and $\c$ are the usual trace anomaly coefficients
for a $d=4$ CFT. 
For field theories admitting a large 
$N$ gravity dual in type IIB supergravity, 
to leading 
order in the $N\rightarrow\infty$ limit 
one has $\a=\c=\pi^2/\kappa_5^2$, where $\kappa_5^2$ is the five-dimensional 
effective gravity constant and we have set the AdS radius to 1.
 In this limit (\ref{Esusyb1b2}) reduces to
\bea\label{EsusylargeN}
\Esusy &=& \frac{(\bone+\btwo)^3}{\bone\btwo}\frac{2\pi^2}{27\kappa_5^2}~.
\eea
In particular the conformally flat $S^1_\beta\times S^3$, 
where $M_3\cong S^3$ is equipped with the standard round metric of radius $r_3$, has $b_1=b_2=1/r_3$, leading to 
$\Esusy=16\pi^2/27r_3\kappa_5^2$. We will 
 reproduce 
(\ref{EsusylargeN}) from a dual supergravity calculation.

\section{II. Dual supergravity solutions}\label{sec2}

The gravity duals are constructed in $d=5$ 
minimal gauged supergravity, whose solutions uplift to type IIB supergravity. 
In Euclidean signature, the bosonic part of the action reads
\bea
S_{\mathrm{bulk}} & =  & -\frac{1}{2\kappa_5^2}\int_{M_5} \Big[ \diff^5 x  \sqrt{\det G}\,   \left(R_{G} - \F_{\mu\nu}\F^{\mu\nu}+12\right)\nn\\
&&   \quad \quad \qquad \qquad -  \tfrac{8\ii}{3\sqrt{3}}\A\wedge \F \wedge \F\Big]~.
\eea
Here $G=(G_{\mu\nu})$ denotes the five-dimensional metric, $R_G$ is its Ricci scalar, $\A$ is the graviphoton and $\F=\diff\A$.

We are interested in supersymmetric solutions that are asymptotically locally Anti-de Sitter (AlAdS), with metric and graviphoton on the conformal boundary given by (\ref{M4}) and (\ref{A}).
Employing a coordinate system defined canonically  by  supersymmetry,
we have solved the supersymmetry conditions and equations of motion  
in a series expansion near the boundary. 
We have then cast the solution in Fefferman-Graham coordinates 
\cite{deHaro:2000vlm}, where the metric is $G  = \diff \tr^2/{\tr^2} +  h_{ij}(\tx,\tr) \diff \tx^i\diff \tx^j$,
and 
\bea
\label{FGexpansion}
h  & = & \frac{1}{\tr^2} \left[h^{(0)} + h^{(2)} \tr^2 + \left(h^{(4)} + \tilde h^{(4)}\log \tr^2\right)\tr^4 + \mathcal{O}(\tr^5)\right],\nn\\
\A & = & A^{(0)} + \left(A^{(2)} + \tilde A^{(2)}\log \tr^2\right)\tr^2 + \mathcal{O}(\tr^3) \ .
\eea
Here the conformal boundary is at $\rho=0$.
The terms at leading order in the expansions, $h^{(0)}\equiv g_4$ and $A^{(0)}\equiv -A/\sqrt{3}$,  
 coincide with the metric (\ref{M4}) and gauge field (\ref{A}), respectively. These depend only on the functions $\wzero(z,\bar{z})$ and $\wone(z,\bar{z})$, which we therefore refer to as \emph{boundary functions}.  $h^{(2)}$,   $\tilde h^{(4)}$, and $\tilde A^{(2)}$ are uniquely fixed in terms of these, whereas $h^{(4)}$ and $A^{(2)}$ are not determined by the conformal boundary, and parametrize the one-point functions of the dual field theories.  These depend on four new functions  $ \newFtwo(z,\bar{z})$, $\newhtwo(z,\bar{z})$,  $ \newhfour(z,\bar{z})$ and $\newhsix(z,\bar{z})$, that we refer to as \emph{non-boundary functions}. The first three of these appear in the expansion of the gauge field:
\begin{widetext}
\bea\label{exprA2tildeA2}
A^{(2)}&=& \tfrac{1}{64\sqrt 3}\Big[\left(-96 \newFtwo - 32\wone  \newhtwo + 4\wone\square \wzero + \tfrac{3}{2}\wone^3 \right)\ii \diff \tau   +  \frac{1}{\wone}\Big( 128 \newhfour -32 \wone\newFtwo -\tfrac{64}{3}\newhtwo^2 + 16\square \newhtwo - \tfrac{32}{3}\newhtwo\square \wzero -16 \wone^2 \newhtwo \nn\\
&&\qquad + \left. 3 \square (\square \wzero + \wone^2) -2(\square \wzero)^2 - \tfrac{5}{3}\wone^2\square \wzero - 3\me^{-\wzero} \partial_z \wone\partial_{\bar z}\wone - \tfrac{5}{12} \wone^4 \Big)\left(-\ii \diff \tau + \diff \psi +a \right)    - *_2 \diff \left(32  \newhtwo + \wone^2  \right)\right] ,\\
\tilde A^{(2)}&=& \tfrac{1}{32\sqrt 3}\left[ \square \wone\,\ii\diff \tau + \big( 2\square \wone - \wone \square \wzero - \tfrac{1}{2}\wone^3\big)(\diff \psi + a) + *_2 \diff \left( 2\square \wzero + \wone^2 \right) \right]~,\nn
\eea
\end{widetext}
where $\Deltanew\equiv  \ex^{-\wzero}\partial_z\partial_{\bar z}$ and $*_2\diff \equiv \ii (\diff\bar{z}\, \partial_{\bar{z}}-\diff z\, \partial_z)$.
A more exhaustive discussion will be presented in \cite{long}.

The bulk action evaluated on a solution is divergent and must be renormalized by the addition of counterterms. As usual, we include the Gibbons-Hawking term
\bea
S_{\mathrm{GH}} &=& -\frac{1}{\kappa^2_5}\int_{\partial M_{\rhoc}} \dd^4 x \sqrt{\det h}\,  K ~,
\eea
to have a well-defined variational principle. Here $h$ is the metric \eqref{FGexpansion} induced on a four-dimensional hypersurface  $\partial M_{\rhoc} = \{\rho=\rhoc=\mathrm{constant}\}$, and $K$ the trace of its second fundamental form. 
The counterterms
\bea
S_{\rm ct} & =  & \frac{1}{\kappa_5^2}\int_{\partial M_{\rhoc}} \!\!\diff^4x \sqrt{\det h} \left(3 + \tfrac{1}{4} R_h \right)\, ,
\label{Counterterm}
\eea
 cancel all divergences as $\rhoc\to 0$.  
In general there is also a $\log \rhoc$ divergence in the action, related to the field theory Weyl anomaly; but in the limit $\epsilon\to 0$ for this class of backgrounds
we have $\mathcal E\equiv 0$, $C_{ijkl}C^{ijkl}\equiv 8\mathcal{F}_{ij}\mathcal{F}^{ij}$, and this term vanishes identically \cite{Cassani:2013dba}. 
Here $R_h$,  $C_{ijkl}$ and  $\mathcal E $ are the Ricci scalar, Weyl tensor and Euler density 
of the metric $h$, respectively. 
 We also include a linear combination of the standard finite counterterms
\bea
\Delta S_\mathrm{st}   & = &  - \frac{1}{\kappa_5^2} \int_{\partial M_{\rhoc}}\!\!\! \dd^4 x \sqrt{\det h} \left( \gammanew\, R_h^2  - \deltanew\, \F_{ij}\F^{ij} \right) , \ \ 
\label{ambij}
\eea
where $\gammanew$ and $\deltanew$ are arbitrary constants. These affect the holographic one-point functions, as well as the on-shell action.
The ordinary renormalized action is obtained as 
\bea\label{eq:Stot}
\Stot \,& \equiv &\, \lim_{\epsilon \to 0} \left(S_{\mathrm{bulk}}+ S_{\rm GH} + S_{\rm ct} + \Delta S_\mathrm{st} \right) ~.
\eea

A variation of the total on-shell action with respect to boundary data takes the form
\bea
\delta \Stot & = & \int_{M_4}   \!\!\diff^4x \sqrt{\det g_4}  \left( -\tfrac{1}{2} T_{ij}  \delta g_4^{ij}  +  j^i   \delta A_i \right)\, ,   \ \ \ 
\label{genvariation}
\eea
where $g_4$ is the $\rho$-independent metric (\ref{M4}) on the conformal boundary.
 The holographic energy-momentum tensor 
$ T_{ij}$   and R-symmetry current $j^i$ may be computed with standard formulas (see \emph{e.g.}\ \cite{Cassani:2014zwa}).
The former is particularly unwieldy, but we have verified  that these satisfy the expected Ward identities.
In particular, the R-symmetry current is conserved, $\nabla_i j^i =0$,  and the energy-momentum tensor obeys the correct conservation equation
\bea
\nabla^i  T_{ij} & = & j^i F_{ji}   ~,
\label{modifiedward}
\eea
with $F =\diff A$.
However, we will  show next that imposing supersymmetric Ward identities requires a non-standard modification of the holographic renormalization scheme.

\section{III. Supersymmetric holographic renormalization}\label{sec3}

According to the gauge/gravity duality, the renormalized on-shell gravitational action is identified with minus the
logarithm of the  partition function of the dual field theory, in the large $N$ limit. Namely
\bea\label{Zmatch}
Z^{\mathrm{susy}}_{S^1_\beta\times M_3} \,& = &\, \mathrm{e}^{- \Stot [M_5]}~,
\eea
where $\Stot [M_5]$ is evaluated on an appropriate supergravity solution, as described in the previous section. Assuming (\ref{Zmatch}), the field theory results summarized
 in the first section imply that $\Stot$ should be invariant under deformations of the boundary geometry that leave fixed
 the transversely holomorphic foliation generated by $\partial_\psi$. Concretely, this implies that 
 $\Stot$ should be invariant under $\wzero\rightarrow\wzero + \deltawzero$, $\wone\rightarrow\wone+ \deltawone$, 
 where $\deltawzero(z,\bar{z})$, $\deltawone(z,\bar{z})$ are arbitrary smooth global functions on $M_3$, invariant under $\partial_\psi$. 
The corresponding variation of $\Stot$ may be computed explicitly using the general  formula (\ref{genvariation}). \hbox{We find}
  \begin{widetext}
\begin{eqnarray}\label{deltaS}
\delta_\wzero \Stot &=& \int_{M_4}\frac{\diff^4 x \sqrt{\det g_4}}{2^63\kappa_5^2}\, \deltawzero\,\big[(1-96\gammanew + 16\tildedelta ) \wone^2 R_{2d}  + \tfrac{1}{2}(1-96\gammanew +28\tildedelta) \Deltanew \wone^2   + 12\tildedelta  \wone\Deltanew \wone \nn\\
\!&&\! \qquad\quad -  \tfrac{1}{32}(19-288\gammanew+192\tildedelta) \wone^4    - 8(-24\gammanew+\tildedelta) (R_{2d}^2 + 2 \Deltanew R_{2d})+ \tfrac{8}{9}\cpsi ( 2\wone R_{2d} + 2\Deltanew \wone - \wone^3) 
 \big]\, ,\\
  \delta_\wone \Stot &=& \int_{M_4}\! \frac{\diff^4 x\sqrt{\det g_4}}{2^{9} 3^2\kappa_5^2} \,
 \deltawone \big[\!\! -24(1-96\gammanew + 16\tildedelta)\wone R_{2d}- 288\tildedelta \Deltanew \wone+ (19-288\gammanew+192\tildedelta) \wone^3+\tfrac{32}{3}\cpsi (3 \wone^2 - 4R_{2d}) \big]\,.\nn
\end{eqnarray} 
 \end{widetext} 
Here $R_{2d}\equiv -\Deltanew \wzero$ is the Ricci scalar of the transverse 
two-dimensional metric $4\ex^{\wzero}\diff z\diff\bar{z}$. We emphasize that this is locally, but not globally, a total derivative. Notice 
the dependence on the constant $\cpsi$, which appears in the boundary gauge field $A$ in (\ref{A}).
In the first variation in (\ref{deltaS})  we hold $\diff\phizero$ fixed, meaning that $\delta(\wone\, \ex^{\wzero})=0$ and hence 
$\deltawone=-\wone\, \deltawzero$; while the second variation in (\ref{deltaS}) is the change in $\Stot$ under an arbitrary variation $\deltawone$.
In obtaining these expressions we have used Stokes' theorem to discard total derivative terms. In particular,
we find that all dependence on the non-boundary functions drops out of these integrals, as does $\diff\lambda(z,\bar{z})$ in (\ref{A}).

Crucially we see that there is no choice of $\gammanew$, $\tildedelta$ for which these 
variations are zero for an arbitrary background. The standard holographic renormalization of the 
previous section hence does not correspond to the supersymmetric renormalization scheme 
used in field theory. This result explains why previous attempts to obtain the 
holographic supersymmetric Casimir energy have failed.

Remarkably, we have found that if we define the new ``finite counterterms''
\bea\label{newCttrms}
\Delta S_{\rm new} & = & -\frac{1}{\kappa_5^2}\int_{M_4}(\ii A\wedge \Phi + \Psi)~,
\eea
where
\bea
\Phi  &\equiv  &  \tfrac{1}{2^33^3} \left(\wone^3-4 \wone R_{2d} \right)\ii \, \ex^{\wzero}\diff z\wedge\diff\bar{z}\wedge (2\, \diff \psi + \ii\, \diff \tau )\ ,\nn\\
\Psi &\equiv & \tfrac{1}{2^{11}3^2}\left( 19 \wone^4 - 48 \wone^2 R_{2d} \right)\diff^4x \sqrt{\det g_4} \, ,
\eea
then (\ref{deltaS}) implies that
\bea\label{Snew}
\Ssusy \,&\equiv &\, \lim_{\epsilon \to 0} \left(S_{\mathrm{bulk}} + S_{\rm GH}+ S_{\rm ct}   \right) + \Delta S_\mathrm{\rm new} 
\eea
 is invariant under  $\wzero\rightarrow\wzero + \deltawzero$, $\wone\rightarrow\wone+ \deltawone$. 
We claim that (\ref{Snew}) is the correct renormalized supergravity action for the class of backgrounds introduced 
in the first section, in the sense that this corresponds to the unique supersymmetric renormalization scheme used
in field theory. In particular, this result should be valid for arbitrary topology of $M_3$. 
 Specializing to the case $M_3\cong S^3$, in the next section we shall not only show that (\ref{Snew}) correctly reproduces (\ref{EsusylargeN}), but moreover we are able to determine the holographic charges in this scheme, and prove that these satisfy the correct BPS relation in field theory.

\section{IV. On-shell action and holographic charges}\label{sec4}

In general to evaluate the bulk action one needs to know 
the full solution. However, with some additional topological assumptions, and assuming 
that a bulk filling exists, one can compute $\Ssusy$ in (\ref{Snew}) explicitly.

We henceforth take $M_3\cong S^3$. In this case the boundary supercharges are 
sections of a trivial bundle, and correspondingly $A$ in (\ref{A}) is a \emph{global} one-form. 
As shown in \cite{Assel:2014paa} this fixes the constant $\cpsi=(\bone+\btwo)/2$, which physically is the 
charge of the spinors under $\partial_\psi$. 
As in the solution of \cite{Cassani:2014zwa}, we assume the  bulk filling is smooth with topology $S^1\times \R^4$, with the bulk graviphoton $\A$ smoothly 
extending $A$ on the boundary. These assumptions, together with supersymmetry, allow one 
to write the bulk action as a total derivative, and hence express $\Ssusy$ as the limit of a term 
evaluated near the conformal boundary. However, this expression still depends on non-boundary functions, 
which are only determined by regularity in the deep interior of the solution. 
Fortunately, we may bypass this problem using another idea from \cite{Cassani:2014zwa}. 
If $\C\cong\R^4$ is a regular hypersurface at  $\tau=$ constant, 
with boundary $M_3\cong S^3$ at infinity, 
then combining the Maxwell equation and Stokes' theorem on $\C$ one can show that 
\bea\label{globalconstraint}
 \int_{M_3}\left(*_5\F + \tfrac{2\ii}{\sqrt{3}}\A\wedge\F\right) \ = \ 0~.
\eea
Substituting \eqref{FGexpansion} and \eqref{exprA2tildeA2} in, this identity may be used to eliminate \emph{all} dependence 
of the on-shell action on non-boundary functions. Also discarding terms which are total derivatives on $M_4$, and noting that \eqref{newCttrms} leads to extensive cancellations, 
\eqref{Snew} evaluates to the remarkable formula
\bea\label{remarkable}
\Ssusy \,&=&\, \frac{\cpsi^2}{27\kappa_5^2}\int_{M_4}\diff^4 x \sqrt{\det g_4}\, R_{2d}~.
\eea
We reiterate that this has been derived here for $M_4\cong S^1_\beta\times S^3$, although 
as we shall explain in \cite{long} this formula has larger validity. As remarked earlier, 
$R_{2d}$ is locally but not globally a total derivative. Its integral is a topological 
invariant of the foliation, proportional to the transverse first Chern class. Using the 
explicit formulas in \cite{Assel:2014paa} for the metric functions and coordinate ranges
for $M_3\cong S^3$ with $\partial_\psi=\bone\partial_{\varphi_1}+\btwo\partial_{\varphi_2}$, we find
\bea
\int_{M_3} \diff^3 x \sqrt{\det g_3}\, R_{2d} &=& 2(2\pi)^2\frac{\bone+\btwo}{\bone\btwo}~.
\eea
Substituting this into (\ref{remarkable}), using $\cpsi=(\bone+\btwo)/2$ and that $\tau$ has period $\beta$, 
we find that $\Ssusy=\beta \Esusy$, where $\Esusy$ is the field theory result (\ref{EsusylargeN})!

The above argument applies to any solution with topology $S^1\times \R^4$, but it is worth emphasizing 
that there are explicit examples. The new counterterms (\ref{newCttrms}) are non-zero even 
for AdS$_5$ in global coordinates, whose boundary is the conformally flat $S^1_\beta\times S^3$ 
geometry with $\bone=\btwo=1/r_3$ mentioned at the end of the first section. The solution 
of \cite{Cassani:2014zwa} has a squashed $S^1_\beta\times S^3_v$ boundary, with the bulk 
solution depending non-trivially on the squashing parameter $v$. However, 
 $\bone=\btwo=1/vr_3$, and we find that $\Ssusy$ is a simple rescaling of the action of AdS$_5$.

Finally, we turn to the holographic charges. Let us start from the standard charges, which may be obtained from $T_{ij}$ and $j^i$ (defined through \eqref{eq:Stot}, \eqref{genvariation}). Due to the Ward identity 
(\ref{modifiedward}) the  canonical Hamiltonian $H$ and angular momentum $J$
associated to translations along $\partial_\tau$ and $-\partial_\psi$ are defined as
\bea
H \,& \equiv  &\,  \int_{M_3} \diff^3x \sqrt{\det g_3} \left( T_{\tau\tau} + j_\tau A_\tau \right)~, \nonumber\\ 
J \,& \equiv &\, \ii \int_{M_3} \diff^3x \sqrt{\det g_3} \left( T_{\tau\psi} +  j_\tau A_\psi \right) ~,
\label{JH}
\eea
respectively.  On the other hand, the holographic R-charge is  defined as 
\bea
Q  \,& \equiv  &\, -\ii \int_{M_3} \diff^3x \sqrt{\det g_3} \, j^\tau ~.
\label{Qele}
\eea
In the dual field theory, these are identified with the vev of the corresponding operators $\langle H \rangle$, $\langle J \rangle$, and  $\langle Q \rangle$.

Utilizing a trick introduced in \cite{Cassani:2014zwa} and elaborated in \cite{long}, one can then show that 
\bea
\beta  H  \ = \  \Stot \qquad \mathrm{and} \qquad  \quad  J  \ = \ 0 ~.
\label{chains}
\eea

Recall that the supersymmetry algebra implies that in the field theory vacuum the BPS relation 
\bea
 \langle  H \rangle  +  \langle  J \rangle +\cpsi \langle Q \rangle \,& = &\, 0~,
\label{BPS}
\eea
should hold, with  $\langle  H \rangle = \Esusy $ \cite{Assel:2015nca}. However, for the Euclidean  AdS$_5$ solution, which 
is expected to correspond to the vacuum of theories in conformally flat space, one
finds that $J  \, |_{\mathrm{EAdS}_5} =   Q  \, |_{\mathrm{EAdS}_5}  =  0$, implying that (\ref{BPS}) is violated.

Assuming that the identity (\ref{genvariation}) holds replacing $\Stot $ with $S_\mathrm{susy}$,
and correspondingly $T_{ij} \to T^\mathrm{susy}_{ij}$, $j_i\to j^\mathrm{susy}_{i}$, we can define
``supersymmetric'' versions of the holographic charges, via formulas analogous to (\ref{JH}) and (\ref{Qele}).
In particular,  the improved  electric charge may be defined as 
\bea
Q_\mathrm{susy} & \equiv & -\ii \int_{M_3}\!\! \frac{\delta S_\mathrm{susy} }{\delta  A_\tau}  =   Q - \frac{1}{\kappa_5^2} \int_{M_3} \!\!\Phi\ ,\quad
\eea
and by direct computation we find
\bea
\cpsi  Q_\mathrm{susy} & = & -\frac{1}{\beta}S_\mathrm{susy}~.
\eea
Moreover, using the relations (\ref{chains}) applied to the improved Hamiltonian and angular momentum,  we deduce that
 $\beta  H_\mathrm{susy}   =   S_\mathrm{susy}$ and $ J_\mathrm{susy}   =  0$, thus showing that these 
 obey the BPS relation (\ref{BPS}).

\section{V. Concluding remarks}

We have constructed new boundary terms of five-dimensional minimal 
gauged supergravity that we argued are necessary to restore supersymmetry of the 
gravitational action in a large class of AlAdS$_5$ solutions. Including 
these counterterms, we have reproduced the supersymmetric Casimir energy and the field theory BPS relation between charges \cite{Assel:2014paa,Assel:2015nca}.   More 
 details, as well as a number of generalizations, will be presented in  \cite{long}. For example, we will perform an analogous computation in four-dimensional
 gauged supergravity, finding that no new counterterms are needed. In  five dimensions we will  consider $M_3$ with more  general topology, making contact with  \cite{Martelli:2015kuk}, as well as a twisting of $S^1_\beta$ over $M_3$.

\bigskip

\noindent P.B.G. is supported by EPSRC and a Scatcherd Scholarship. D.C. is  supported  by  the  European  Commission  Marie  Curie  Fellowship PIEF-GA-2013-627243.  D.M. acknowledges support from ERC Starting Grant N. 304806.

\end{document}